\documentstyle[amssymb,aps,prl,multicol,epsfig]{revtex}

\newcommand{\beq}{\begin{equation}}
\newcommand{\eeq}{\end{equation}}
\newcommand{\beqa}{\begin{eqnarray}}
\newcommand{\eeqa}{\end{eqnarray}}
\newcommand{\non}{\nonumber}
\newcommand{\xiv}{{\vec{\xi}}}
\newcommand{\Rv}{{\vec{R}}}
\newcommand{\Ocal}{{\cal O}}

\newcommand{\av}{{\vec{a}}}
\newcommand{\omegav}{{\vec{\omega}}}
\newcommand{\etav}{{\vec{I}}}
\newcommand\mean[1]{\langle#1\rangle}
\newcommand\mmean[1]
        {\langle\hspace{-0.06cm}\langle#1\rangle\hspace{-0.06cm}\rangle}

\newcommand\bp{{\mathbf p}}
\newcommand\bs{{\mathbf s}}
\newcommand\bW{{\mathbf W}}
\newcommand\bx{\nabla_{\mathbf s} {\cal H}}
\newcommand\bm{{\mathbb M}}

\begin{document}
\draft

\title{Continuous time dynamics of the Thermal Minority Game}

\author{
Juan P. Garrahan\cite{thanksJPG},
Esteban Moro\cite{thanksEM}
and 
David Sherrington\cite{thanksDS}
}

\address{
Theoretical Physics, University of Oxford,
1 Keble Road, Oxford OX1 3NP, United Kingdom
}

\date{April 17, 2000}

\maketitle

\begin{abstract}
We study the continuous time dynamics of the Thermal Minority Game. 
We find that the dynamical equations of the model reduce to a set of
stochastic differential equations for an interacting disordered system 
with non-trivial random diffusion. This is the simplest 
microscopic description which accounts for all the 
features of the system.
Within this framework, we study the phase structure of the model and
find that its macroscopic properties strongly depend on the 
initial conditions.
\end{abstract}

\begin{multicols}{2}
\narrowtext

Many of the current challenges for statistical 
physics have their origins in problems in
biology \cite{bio} 
and economics \cite{pwa,doyne}. In particular,
the application of 
ideas and techniques of the statistical mechanics of
disordered systems
can prove useful in 
the study of systems of
adaptive and competitive agents, which are relevant, for example,
to the microscopic modeling of financial markets; 
and, conversely,
such problems can raise new issues for statistical physics.
One of these systems is 
the minority game (MG) \cite{cz,savit}, a simple
model based on Arthur's ``El Farol'' bar
problem \cite{arthur} 
which describes the behaviour of a 
group of competing heterogeneous agents subject to the economic 
law of supply and demand. Despite its simplicity, the MG is very
non-trivial, and although much progress has been made in the
qualitative \cite{neil,rene,neil2} 
and quantitative \cite{cm,cmz} understanding
of its features, a full analytic solution of the MG
is still missing.

The main hurdles in the way of an analytical study of the MG in its
original formulation were its non-locality in time due to the
dependence on the game history,
its discrete kinematics and dynamics, and the ``best-strategy'' rule
(see, however, \cite{cm}).
The first of these obstacles was overcome 
in \cite{andrea}, where it was shown numerically that
the macroscopic behaviour of the MG was unchanged if 
the real history was replaced by a random one.
This allowed the study of a simpler stochastic Markovian problem
instead of the original deterministic non-Markovian one. 

In \cite{mino} a natural 
continuous generalization of the MG was presented.
The ``information'' to which the agents reacted was taken 
as an external input to the system and it
was shown that all the macroscopic features of the MG were
preserved, as long as the external information was ergodic in time,
the simplest choice being just noise.
To handle the problem of the `best-strategy' rule, the Thermal
Minority Game (TMG) was introduced, in which a certain degree of
stochasticity in the choice of the strategies by the agents
was allowed, controlled
by a parameter $T$, the ``temperature'', 
the limits $T=0$ and $T=\infty$ corresponding 
to the original deterministic 
MG and the case of completely random strategy 
choices, respectively.
The TMG displayed extra non-trivial structure as a function of $T$,
notably that in the region where the MG performs worse than random,
the system can be made to perform better than random 
by allowing a certain degree of 
individual stochastic error.

In the present paper we carefully study the continuous
time limit of the TMG, in order to obtain the simplest microscopic 
description which accounts for {\em all} the macroscopic features of 
the system, and 
as a further step towards an analytical solution of the model.
We confirm that
the external information `integrates-out' providing simply 
an effective coupling between agents.  
We also show the crucial dependence of the macroscopic properties of 
the model on the initial conditions.
We find that the microscopic equations of the TMG 
reduce to a set of disordered stochastic differential equations 
with a non-trivial random diffusion matrix, 
and study the phase structure of the model in the
temperature/dimension plane. 

The setup of the TMG is as follows \cite{mino}. 
The system consists of $N$ agents playing the game. 
At each time step $t$, 
each agent 
reacts to a common piece of ``information'' $\etav(t)$, 
by making a ``bid'' $b_i(t)$ ($i=1,\dots,N$). 
The information, 
defined as 
a unit-length vector in ${\mathbb R}^D$, is taken to 
be a random noise, $\delta$-correlated in time
and uniformly distributed on the unit sphere \cite{any}.
The bid $b(t)$ is defined to be a real
number, which can be interpreted as placing an order in a market, 
of size $|b(t)|$ and
positive/negative meaning buy/sell.
Bids are prescribed by ``strategies'': maps from information to bid, 
${\mathbb R}^D \to {\mathbb R}$.
For simplicity the strategy space $\Gamma$
of the model is restricted to the 
subspace of homogeneous linear mappings.
A strategy $\vec R$ is defined
as a vector in 
${\mathbb R}^D$, subject to the constraint $\| \vec R \| =
\sqrt{D}$, and the prescribed bid is 
just the scalar product
$\vec R \cdot \etav(t)$.
Each agent has $S$ strategies, drawn randomly and independently from 
$\Gamma$ (with uniform distribution) remaining fixed throughout
the game.
In what follows we will restrict for simplicity 
to the case of two strategies per agent $S=2$,
the generalization to $S > 2$ being straightforward. 
At time step $t$ each agent $i$ chooses one of his/her strategies 
$\vec R^\star_i(t)$ to play with. 
The ``total bid'' (or ``excess demand'') is then
$A(t) \equiv \sum_{j} b_j(t) 
        = \sum_{j} \vec R^\star_j(t)\cdot \etav(t)$. 
The agents keep track of the potential success of the strategies 
by assigning points to them, which are updated 
according to 
$P(\vec R,t+1) = P(\vec R,t) - A(t) \, b(\vec R) / N$, 
where $P(\vec R,t)$
represents the points of strategy $\vec R$
at time $t$.

In the original formulation of the MG the agents played in a
deterministic fashion using their
`best' strategies, the ones with the highest number of points.
In the TMG the natural
generalization to non-deterministic behaviour is
allowed.
At time step $t$, each agent $i$ chooses
$\vec R^*_i(t)$ randomly from his/her
$\{ \vec R_i^1, \vec R_i^2 \}$ with
probabilities $\{ \pi_i^1(t), \pi_i^2(t) \}$. 
The probabilities $\pi^a_i(t)$ are functions of the points
parameterized by a temperature $T$, 
defined so as to interpolate between 
the MG case at $T=0$, 
all the way up to the totally random case
$\pi_i^1=\pi_i^2=1/2$ at $T=\infty$. 
The qualitative behaviour of the system does not depend on the 
specific functional form of the $\pi^a_i(t)$. 
In \cite{mino}
the probabilities were defined as 
$\pi^{1,2}_i(t) \propto \exp{[\beta P(\vec R_i^{1,2},t)]}$
(with $\pi^1_i(t)+\pi^2_i(t)=1$ and $\beta=1/T$), 
while an alternative convenient form is given by
$\pi^{1,2}_i(t) \propto \exp{[\pm \beta z_i(t)]}$, 
where $z_i(t) \equiv {\mathrm sgn}(p_i(t))$, 
with $p_i(t) \equiv [P(\Rv_i^1,t)-P(\Rv_i^2,t)]/2$. 
The consequential difference between these two definitions 
will be discussed below. 

The set of unconstrained degrees of freedom of the TMG
is given by the difference $p_i(t)$
of the points of the two
strategies of each agent.
The choice of strategies can then be defined by
$\vec R^*_i(t) = \vec h_i + \vec \xi_i \,   
    {\rm sgn} \left[ s_i(t) + \mu(t) \right]$,
where  
$s_i(t) \equiv \pi_i^1(t) - \pi_i^2(t)$,
$\vec \omega_i \equiv \left( R_i^1 + R_i^2 \right)/2$,  
$\vec \xi_i \equiv \left( R_i^1 - R_i^2 \right)/2$, 
and $\mu(t)$ is a stochastic
random variable uniformly distributed between 
$-1$ and $1$ and independently distributed in time.
The equations for the point
differences then read,
\beq
p_i(t+1) = p_i(t) - \av(t) \cdot \etav(t) \, 
    \vec \xi_i \cdot \etav(t)
    \label{pipi} ,
\eeq
where $\av(t) \equiv \sum_{i} \vec R^\star_i(t)/N$. 
Eqs.\ (\ref{pipi}),  
together with the random processes for 
$\etav(t)$ and $R_i^*(t)$, define 
the dynamics of the TMG.

The average of the total bid $A(t)$
over time and quenched disorder is zero,
so the first relevant macroscopic 
observable of the TMG is its normalized 
standard deviation $\sigma$ (or ``volatility'')
$\sigma^2 \equiv
    N^{-1} \overline{\langle A^2(t) \rangle}$,
where the overline means disorder average,
and $\langle \cdot \rangle  \equiv \lim_{t \to \infty}
\frac{1}{t}\int_{t_0}^{t_0+t} (\cdot) \,  dt'$.
In \cite{mino} it was found that $\sigma$ had a non-trivial structure
both as a function of the reduced dimension
of the strategy space $d=D/N$ and of the temperature.
The second important observable is the fraction
$\phi$ of ``frozen'' agents, 
defined as those for which one of the strategies 
always outperforms the other, 
$\phi \equiv N^{-1} \sum_i
    \delta\left(\left| \langle z_i(t) \rangle
    \right| - 1\right)$, with the normalization $\delta(0)=1$.
It was introduced in \cite{cm} as an order parameter for the MG,
where it was found that $\phi(d)$ changed from zero to a finite value
at $d=d_c$.

We now consider the continuous time limit of
Eqs.\ (\ref{pipi}) in such a way as to preserve all the macroscopic 
features of the TMG. 
To this end we introduce an arbitrary time step $\Delta t$.
We deal first with the information $\etav(t)$. 
Let us assume that $\etav(t)$ is a differential random motion
in the space of strategies, i.e.,
$\etav(t) = \Delta \vec W(t)$,
with zero mean and variance $\Delta t$. 
Replacing in Eqs.\ (\ref{pipi}) we obtain
$p_i(t+\Delta t) = p_i(t) - 
    \av(t) \cdot \Delta \vec W(t) \,
    \xiv_i \cdot \Delta \vec W(t)$. In the
limit $\Delta t \to 0$, and using the 
Kramers-Moyal expansion \cite{vankampen}, 
we get
\beq\label{eqcont_1}
d p_i(t) = -    
    \av(t) \cdot \xiv_i \, dt/D 
    + \Ocal(dt^2) .
\eeq
Note that to $\Ocal(dt)$ the noise has been eliminated
in favour of an effective interaction among the agents, and
the $\sigma$ becomes 
$\sigma^2 = (ND)^{-1} 
\overline{\langle \av(t) \cdot \av(t) \rangle}$.

\begin{figure}
\begin{center}
\epsfig{file=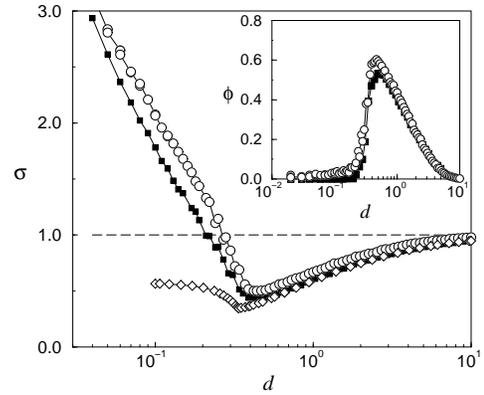, width=2.8in}
\caption{
Volatility $\sigma$
as a function of the reduced dimension $d=D/N$. 
Squares $\blacksquare$ 
correspond to the original dynamics Eqs.\ (\ref{pipi}); 
circles $\bigcirc$
to simulations of Eqs.\ (\ref{eqt0}), 
where an Euler algorithm has been used with time step $dt = 0.05$; 
diamonds $\Diamond$ 
to the approximation of [11].
In the inset we show $\phi$ as a
function of $d$. Average over $100$ samples; $N=100$; $t=t_0=10^4$. 
}
\label{fig1}
\end{center}
\end{figure}

At $T=0$, corresponding to the MG, 
Eqs.\ (\ref{eqcont_1}) are completely
deterministic. To first order in $dt$ we have
\beq
dp_i(t) = - [ h_i + \sum_j J_{ij} \, z_j(t) ] \, dt ,
        \label{eqt0}
\eeq
where 
$h_i \equiv \sum_j \omegav_j \cdot \xiv_i / ND$
and $J_{ij} \equiv \xiv_j\cdot\xiv_i /ND $, 
while the volatility reads
\beq
\sigma^2 = 
        \overline{
        \Omega 
        }
        +
        2 \sum_i 
        \overline{
        h_i \, \mean{z_i(t)} 
        }
        + \sum_{ij} 
        \overline{
        J_{ij} \, \mean{z_i(t) \, z_j(t)}       
        } ,
\eeq 
and 
$\Omega \equiv \sum_{ij} \omegav_i \cdot \omegav_j / ND$.
In order to check the continuous time approximation
at $T=0$ 
we have simulated 
Eqs.\ (\ref{eqt0}). Results are presented in Fig.\ \ref{fig1}.
We can see that this approximation reproduces all the features of
the original MG. Note that in Eqs.\ (\ref{eqt0}) all stochasticity
coming from the information has dropped out, the only effect
being a small quantitative deviation in the low $d$ region.

\vspace{-0.4cm}
\vspace{-0.4cm}

Eqs.\ (\ref{eqt0}) can be rewritten as
$d\bp/dt = - \nabla_{\bs} {\cal H}$, 
where $\bp \equiv (p_1, \ldots, p_N)$, similarly for $\bs$, 
and 
\beq
{\cal H}= \frac{1}{2} \Omega + 
        \sum_i h_i s_i + 
        \frac{1}{2} \sum_{ij} J_{ij} s_i s_j .
\eeq
This is similar to what was done in \cite{cmz} for the time and 
information averages of $z_i$. There
the value of $\sigma$ was related to the
average extrema of ${\cal H}$ 
by assuming that the system equilibrated. A good agreement with the 
numerics was found in
the phase $d>d_c$, but this method failed to reproduce 
the behaviour in the $d<d_c$ 
phase (see Fig.\ \ref{fig1}). 
This disagreement was speculated as due to the need for
terms with higher order time derivatives 
in the continuous time equations.  
This is clearly wrong, 
since, as we have just shown, 
Eqs.\ (\ref{eqt0}) describe correctly the dynamics of the model
for {\em all} values of $d$ 
(see Fig.\ \ref{fig1}).

\begin{figure}
\begin{center}
\epsfig{file=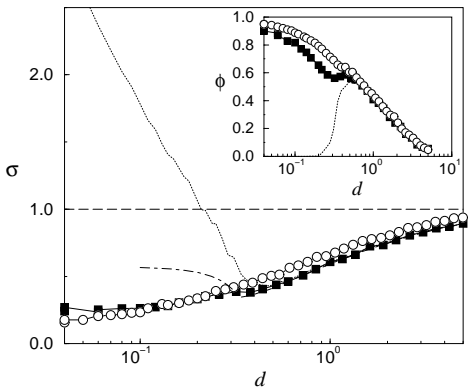, width=2.8in}
\caption{
Volatility 
as a function of $d$ for random initial
conditions $|p_i(0)| = \Ocal(1)$, for the 
original dynamics Eqs.\ (\ref{pipi}) and 
the continuous time approximation 
Eqs.\ (\ref{eqt0}). Dotted lines correspond to
zero initial conditions and
the approximation of [11].
In the inset we show the fraction of frozen agents $\phi$. 
Symbols and details of the simulations are those of the previous
figure.
}
\label{fig2}
\end{center}
\end{figure}

The phase $d<d_c$ of the MG is very sensitive to the initial conditions.
In Fig.\ \ref{fig2} we show the results of simulating both the original
dynamics Eqs.\ (\ref{pipi}) and the continuous time approximation 
Eqs.\ (\ref{eqt0}) starting from random initial conditions $|p_i(0)| =
\Ocal(1)$. 
From Fig.\ \ref{fig2} see that 
the behaviour of both
$\sigma$ and $\phi$ is different from that of
Fig.\ \ref{fig1} in the region $d<d_c$:
the system stays in the better-than-random phase for all values of $d$.
Again the continuous time dynamics is very close to
the original discrete one. 
This sensitivity of the results to the initial conditions
is a clear indication that the system does not equilibrate for
$d<d_c$, and raises a question on the existence and
character of the ``phase transition'' in the MG \cite{cm}. 

\vspace{-0.4cm}
\vspace{-0.4cm}

When the temperature is different from zero
the TMG Eqs.\ (\ref{eqcont_1}) still depend on the
stochastic choice of strategies $R_i^*(t)$, even at leading order. 
At each time step, $R_i^*$ takes one of the two possible values 
$R_i^{1,2}$, defining a stochastic jump process. In order to write
the corresponding Master Equation we need to know the transition
probabilities. 
The r.h.s. of Eq.\ (\ref{eqcont_1}), which we denote $\Delta_i$,
is a normalized sum of $N$ random numbers
$\xiv_i \cdot R_j^*(t)$, each with mean
$\xiv_i \cdot \omegav_j + \xiv_i \cdot \xiv_j \, s_j(t)$,
and variance
$( \xiv_i \cdot \xiv_j )^2 [ 1 - s_j^2(t)]$.
By the central limit theorem, we know that for $N$ large
$\Delta_i$ will tend to be normally distributed with mean
$\mmean{\Delta_i} = \partial {\cal H}/\partial s_i$, 
and  variance 
$\mmean{\Delta_i^2} = \sum_j J_{ij}^2 [ 1 - s_j^2(t)]$, where
$\mmean{\cdot}$ stands for average over realizations of
the random process $\mu(t)$. 
Moreover, $\Delta_i$ and $\Delta_{j \neq i}$ are correlated,
the covariance matrix given by 
\beqa
M_{ij}[\bp(t)] &\equiv& 
        \mmean{\Delta_i \, \Delta_j} - 
        \mmean{\Delta_i}
        \mmean{\Delta_j}
        \non  \\
        &=&
        \sum_k J_{ik} \, J_{jk} \left[ 1 - s_k^2(t) \right] .
        \label{corr}
\eeqa
Collecting these results, we obtain the transition probabilities
in the large $N$ limit, 
$W(\bp'| \bp) =  \Phi(\bx; \bm)$,
where $\Phi$ corresponds to the normal distribution with mean
$\bx$ and covariance 
matrix $\bm \equiv \{ M_{ij} \}$. 
Note that $\partial {\cal H}/\partial s_i \sim \Ocal(1)$, 
and $M_{ij} \sim \Ocal(1/N)$, so that 
fluctuations are also of $\Ocal(1)$ 
and thus are {\em not} suppressed when $N \to \infty$.

The $R_i^*(t)$ are chosen independently at 
each time. If we make the natural assumption 
that in the limit $dt \to 0$ their 
correlation at different times is a $\delta$-function, 
the Master Equation becomes a Fokker-Planck equation 
by means of Kramers-Moyal expansion \cite{vankampen}
\beq
\frac{\partial {\cal P}}{\partial t}
        = 
        - \sum_i \frac{\partial}{\partial p_i}
        \left( \frac{\partial{\cal H}}{\partial s_i} 
        \, {\cal P} \right) 
        + \frac{1}{2} \sum_{ij}
        \frac{\partial^2}{\partial p_i \partial p_j}
        \left( M_{ij} \, {\cal P} \right)  .
\eeq
We therefore conclude that the dynamics of the TMG 
is effectively described by
a set of stochastic differential equations for the point
differences
\beq
d\bp = - \bx \, dt +    
        \bm \cdot d\bW ,
        \label{sde}
\eeq
where $\bW(t)$ is an $N$-dimensional Wiener process, 
and the volatility is given by
$\sigma^2 = 2
        \overline{ \mean{{\cal H}}} 
        + \sum_i \overline{J_{ii}}
        - \sum_i \overline{J_{ii} \mean{s_i^2}}$.
 
\begin{figure}
\begin{center}
\epsfig{file=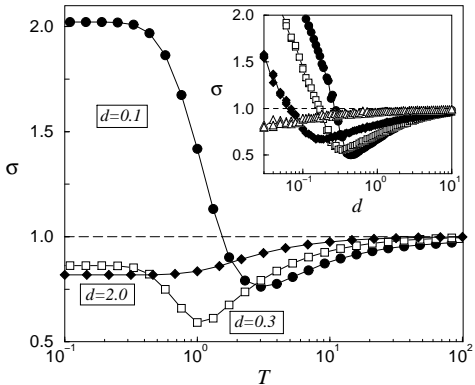, width=2.8in}
\caption{
Volatility 
as a function of the temperature from the continuous dynamics
Eqs.\ (\ref{sde}). Inset: 
volatility as a function of $d$ 
for different values of the 
temperature $T=10^{-3}, \, 1, \, 2, \, 10$ ($\bullet$, 
$\Box$, $\blacklozenge$, and 
$\bigtriangleup$, respectively). 
A second order stochastic Heun algorithm has 
been used with time step $dt = 0.02$. Average over $20$ realizations
of the Wiener process and
$50$ samples; $N=100$; $t=t_0=10^4$; initial conditions $\bp(t=0)=0$.
}
\label{fig3}
\end{center}
\end{figure}

\begin{figure}
\begin{center}
\epsfig{file=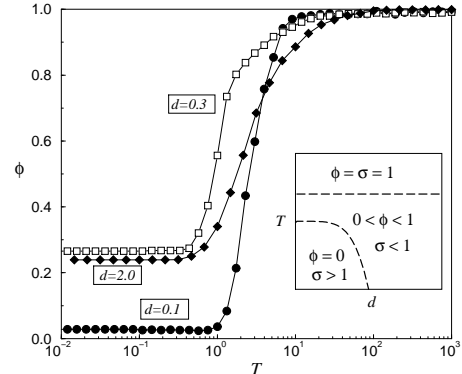, width=2.4in}
\caption
{
Fraction of
frozen agents $\phi$ as a function of $T$. 
Inset: schematic phase
diagram of the TMG in the $(d,T)$ plane. Dashed lines indicate 
crossovers rather than sharp transitions. 
}
\label{fig4}
\end{center}
\end{figure}

We have checked by means of extensive numerical simulations that 
Eqs.\ (\ref{sde}) give the {\em same} 
results as Eqs.\ (\ref{eqcont_1}), up to statistical errors.
Figs. \ref{fig3} and \ref{fig4} present the 
results from the continuous time dynamics Eqs.\ (\ref{sde}).
For these simulations
we have chosen for the strategy-use probabilities the form 
$\pi^{1,2}_i(t) \propto \exp{\pm \beta z_i(t)}$ which makes the
numerics simpler. Similar results can be obtained with the 
form $\pi^{1,2}_i(t) \propto \exp{\beta P(\vec R_i^{1,2},t)}$, 
but a small-$p$ cutoff of ${\cal O}(dt)$ is required 
to avoid the system getting trapped in the $p=0$ region. 
In Fig.\ \ref{fig3} we plot the volatility as a
function of the temperature, showing that 
the behaviour is the same as the one found in
\cite{mino}: for $d<d_c$, as $T$ is increased
$\sigma$ first drops to a minimum, and then increases
towards the random case $\sigma=1$; for $d>d_c$, the optimum value is
the MG one, and $\sigma$ simply grows monotonically to $1$.
In the inset we give $\sigma$ as a function of $d$ for different
temperatures.

Fig.\ \ref{fig4} shows 
how the fraction of frozen agents varies as the temperature is
increased. 
For all values of $d$ there is
a clear jump at $T={\cal O}(1)$ from the MG value to $\phi=1$ 
\cite{frozen}.
Figs.\ \ref{fig3} and \ref{fig4} determine the phase diagram of the 
TMG in the $(d,T)$ plane. It is schematically depicted 
in the inset of Fig.\ \ref{fig4}.
For low $d$ and $T$ the system performs worse than random,
while for large enough values of $T$ the system becomes random,
independently of $d$. Better than random performance is achieved 
between these two regions. It is important to note that, as in the
case of the MG, the phase
structure of the TMG depends strongly on the initial conditions. In
particular, the low $(d,T)$ part of the phase diagram shrinks to zero 
for finite random initial conditions.

\vspace{-0.4cm}
\vspace{-0.4cm}

In the case where the probabilities are defined as 
$\pi^{1,2}_i(t) \propto \exp{\beta P(\vec R_i^{1,2},t)}$, 
the monotonic increase of $\sigma$
to the random value at large $T$ reported in \cite{mino}
is due to 
finite waiting times, 
as pointed out in \cite{comment}:
for $t \gg NT$ the volatility
stays at the minimum value for any finite $T \gg 1$.
This phenomenon is easily understood from Eqs.\ (\ref{sde}). 
For large values of $T$, there is first a transient in which 
$s_i=\tanh{\beta p_i}$
are close to zero, and Eqs.\ (\ref{sde}) reduce to 
$dp_i \approx - h_i \, dt + \sum_{jk} J_{ik} J_{jk} dW_j$, 
i.e., the point differences
of all the agents
do a randomly biased Brownian motion, and the system performs 
as in the random case. Eventually, however, 
$p_i$ become of ${\cal O}(T)$
and $s_i$ finite, and the system effectively behaves as for $T
\sim {\cal O}(1)$. Note that this cannot happen 
when $s_i = \tanh{\beta z_i}$. 

Eqs.\ (\ref{sde}) are much simpler than the original ones for the
microscopic description of the TMG. 
The external information has been replaced by
an interaction among the agents and the random strategy choice has
given rise to the diffusive term.
They describe a dynamics which is different from the 
relaxation of disordered systems usually found
in physical problems: the random force $\bx$ can be written as 
the gradient of a potential function only up to a factor, which
amounts to a metric in the space of $\bp$, and 
the non-trivial diffusion matrix $\bm$
depends both on the variables $\bp$ and on the quenched 
disorder of the problem. Finding adequate analytic
asymptotic solutions
to this dynamics is the next challenging task.

We thank J.-P. Bouchaud, A. Cavagna, 
I. Giardina, G. Lythe 
and D. Williams for useful discussions. 
This work was supported by EC Grant No. ARG/B7-3011/94/27 and 
EPSRC Grant No. GR/M04426.

\vspace{-0.3cm}

\end{multicols}


\begin{thebibliography}{}
\bibitem{any}
Any other normalized isotropic distribution in ${\mathbb R}^D$, e.g.,
a Gaussian, would be equally suitable. The same applies to the
strategies.
\bibitem{frozen}
When $T>0$, the fraction of agents playing always the same
strategy is not given by $\phi$ but 
rather by $|\mean{s}| \propto \phi \, \tanh{\beta}$.
\end{thebibliography}
\end{document}